# PREPARATION OF ODA-CLAY HYBRID FILMS BY LANGMUIR-BLODGETT TECHNIQUE[*]

P. K. PAUL, S. A. HUSSAIN, D. BHATTACHARJEE[†]

*Department of Physics, Tripura University,
Suryamaninagar-799130, Tripura, India*[‡]
*tuphysic@sancharnet.in*[§]
Fax: +91381 2374801 (O)

Hybrid monolayers of clay minerals (hectorite) and Octadecyamine (ODA) were prepared using the Langmuir-Blodgett (LB) technique. Surface pressure-area per molecule isotherm, FTIR and atomic force microscopy were used to confirm and analyze the ODA-hectorite hybrid films. The monolayer thickness is 2 nm and average height, length and width of individual clay platelets ranges between 1.5 to 2 nm, 500 to 1250 nm and 100 to 115 nm respectively. The surface coverage was more than 80%..

## 1. Introduction

Clay minerals play important roles in the modern technology. New materials of functional ceramics are made from mica. [1,2] Organo-clay composites are now being extensively investigated in material science. [3–5] Thin films of clay minerals have been studied in application to modified electrodes, sensors, photochromic devices, nonlinear optical devices, and so on. [6–14] One of the outstanding properties of the clay is the simultaneous incorporation of polar or ionic molecules into the interlamellar spaces (intercalation) [15-17] resulting in hybrid materials. The property of intercalation makes it easy to prepare the composite materials. If the orientation of the incorporated molecules can be controlled, the clay composite materials would be applicable to devices for current rectifying, nonlinear optics, and one-way energy transfer.

There are several methods to fabricate these hybrid films such as spin coating, casting, self-assembly and Langmuir-Blodgett (LB) techniques. [18-20] Layer-by-layer self-assembly (LBL) and LB technique are attractive among these methods because they afford control of the organization at the molecular level. [21] Self-assembly is a simple method to construct layered materials. However, the control over two- and three-dimensional ordering is limited. The LB technique enables fabrication of ordered monolayers and organized molecular assemblies with well-defined molecular orientation onto desired substrates. It offers a tool to realize the ordered assembly of hybrid monolayers.[22]

---

[†] Corresponding author

In Langmuir-Blodgett (LB) method the floating monolayer at the air-liquid interface is deposited onto a solid substrate in a layer-by-layer method to make a multilayer. When amphiphilic cations are spread onto an aqueous clay dispersion of a clay mineral in a LB trough, negatively charged clay platelets in the suspension are adsorbed electrostatically onto the bottom of the floating monolayer of the cations at the air-clay dispersion interface.[23] The hybrid monolayer of the clay platelets and the amphiphilic cations thus formed can be deposited onto a solid substrate to form mono- and multilayer LB films.[24] Thus the hybrid monolayers consist of one layer of inorganic clay particles covered on one side by one layer of adsorbed amphiphilic molecules. By changing various LB parameters it is possible to control the properties of such hybrid films. In principle, such monolayers are very versatile for the following reasons: 1) the inorganic clay particles give mechanical strength and mechanical stability to the films. 2) The amphiphilic molecules can carry a functionality such as: acid-base, redox and catalytic centres.[25] The layer of inorganic particles can act as an insulator and can also contain specific physico-chemical functions such as light absorbance and emission and redox functionality.[26]

In this communication we report the preparation of nano-order organo-clay hybrid films by Langmuir-Blodgett (LB) method. We used ODA as amphiphilic cation and clay platelets hectorite. Pressure-area isotherm, FTIR and AFM observations confirm the formation of organo-clay hybrid films.

**2. Experimental**

2.1. *Materials*

Octadecylammine (ODA, 99%, Sigma Aldrich, USA) was used as received. ODA was dissolved in HPLC grade chloroform (99.9 % Aldrich, stabilized by 0.5-1% ethanol). The clay mineral hectorite used in this study was obtained from the Source Clays Repository of the Clay Minerals Society. The clay minerals was $Na^+$ saturated by repeated exchange with 1 M NaCl solution and washing. The particle size fraction between 0.5 and 2 μm was obtained by centrifugation. The clay mineral was stored as freeze-dried powders.

2.2. *Isotherm measurement and hybrid film formation*

A commercially available Langmuir-Blodgett (LB) film deposition instrument (APEX 2000C, India) was used for isotherm measurement and hybrid monolayer film preparation. Clay dispersions stirred for 24 h (by a magnetic stirrer) in Milli-Q water were used as subphase. The clay concentration was fixed at 10 mg/L. 75 micro-litre of chloroform solution of ODA (0.5 mg/ml) was spread on the air-clay dispersion interface of the LB trough. When the clay particles come in contact with the floating ODA monolayer strong electrostatic interaction is occurred between the negatively charged hectorite and the positively charged ODA. Thus hectorite comes onto the air-clay dispersion interface and a hybrid Langmuir monolayer is formed. A schematic diagram of the hybrid monolayer formation and consequent deposition onto solid substrate is shown in figure 1.

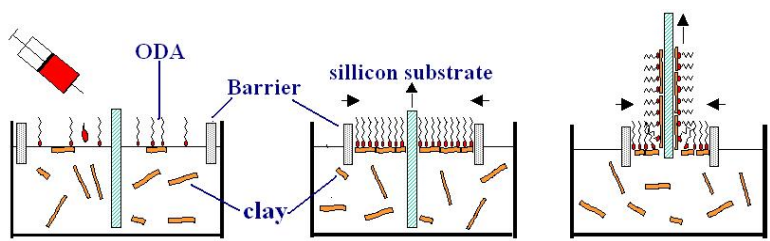

Fig.1. Schematic diagram of the formation of hybrid Langmuir and Langmuir-Blodgett films.

Allowing 30 minutes to complete the hybrid film formation, the floating hybrid monolayer was compressed at a rate of 10 cm$^2$min$^{-1}$ to monitor the pressure-area isotherm. Surface pressure was recorded using a Wilhelmy arrangement. Each isotherm was repeated and consistent results were obtained. Monolayer films were deposited in upstroke (lifting speed 5 mm min$^{-1}$) at a fixed surface pressure of 0 and 15 mN/m onto smooth silicon substrate for AFM measurement and onto a ZnSe substarte for FTIR measurement. The transfer ratio was found to be 0.98 $\pm$ 0.02.

### 2.3. *AFM measurement*

The Atomic force Microscopy (AFM) images of ODA-hectorite hybrid monolayer films were taken in air with a commercial AFM system Autoprobe M5 (Veeco Instr.) using silicon cantilevers with a sharp, high apex ratio tip (UltraLevers$^{TM}$, Veeco Instr.). All the AFM images presented here were obtained in intermittent-contact ("tapping") mode. Typical scan areas were 3×3 μm$^2$ and 0.750×0.750 μm$^2$. The monolayers on Si wafer substrates were used for the AFM measurements.

### 2.4. *FTIR measurement*

ATR-FTIR spectra were obtained with the Bruker IFS66v/s spectrometer on monolayers deposited on ZnSe substrate. Prior to recording the spectra of LB films on ZnSe, the background spectra was recorded using a clean ZnSe substrate. For FTIR measurement of hectorite and ODA in KBr pallet the background spectra was recorded using a pure KBr pallet without any sample.

## 3. Results and Discussions

### 3.1. *Pressure-area isotherm at air-water interface*

The surface pressure – area per molecule (π-A) isotherms of ODA in the absence and presence of a hectorite particles (10 ppm) are given in figure 2. The pure ODA isotherm is a smoothly rising curve with lift-off area 18.5 $A^{0^2}$, which is consistent with the area occupied by densely packed alkyl chains (about 20 $A^{0^2}$). It shows steep rising up to collapse pressure is reached at about 55.7 mN/m. This behaviour is consistent with the previously reported results.[27] On the other hand the ODA isotherm in presence of hectorite is totally different in shape and nature. This is a direct evidence of the incorporation of clay into ODA monolayer and formation of ODA-hectorite clay hybrid films. Here the isotherm started with initial lift-off area of 56.6 $A^{0^2}$ and collapsed at around 50.9 mN/m surface pressure. The isotherm is also less steeper than that in the absence of hectorite. As a whole the hybrid isotherm is shifted to the larger area side in comparison to the ODA isotherm in the absence of clay. This is due to the change in nature and properties of the hybrid films with respect to the pure components.

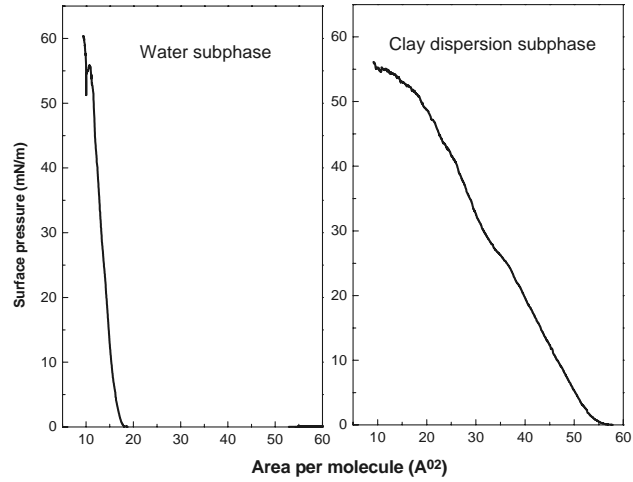

Fig.2. Surface pressure-area per molecule isotherm of ODA in absence and presence of clay (10 ppm hectorite).

Compressibility of the monolayer films was calculated according to the relation;

$$C = -\frac{1}{a_1}\frac{a_2 - a_1}{\pi_2 - \pi_1}$$

where $a_1$ and $a_2$ are the areas per molecules at surface pressures $\pi_1$ and $\pi_2$ respectively.[28] In the present case $\pi_1$ and $\pi_2$ are chosen as 10 and 30 mN/m respectively. It was observed that the compressibility of ODA monolayer in the absence of clay particle is 6.76 mN$^{-1}$ and that in presence of clay particles is 15.9 mN$^{-1}$. Therefore it is interesting that the compressibility of the monolayer films increases with the inclusion of clay particles in the monolayer films. This indicates that incorporation of clay particles makes the films more elastic and compressible. All the observed changes between the isotherms of ODA in absence and presence of clay are indicative of the interaction between ODA and clay mineral platelets with the formation of hybrid monolayer at the air-water interface.

### 3.2. *FTIR spectroscopy*

Figure 3 shows the FTIR spectra of ODA-hectorite clay hybrid monolayer film deposited onto ZnSe substarte at a surface pressure of 15 mN/m along with the spectra of hectorite and ODA in KBr pallet for comparison.

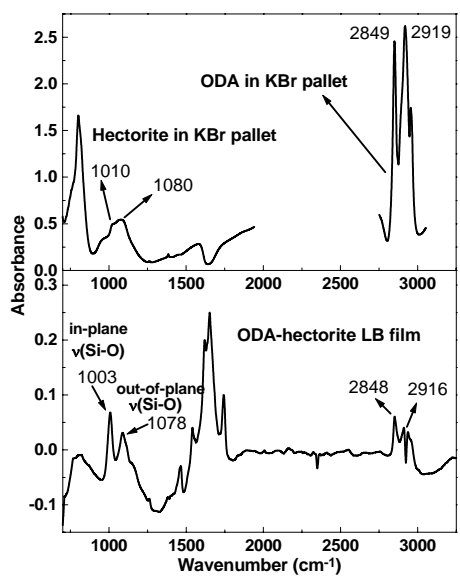

Fig. 3. FTIR spectra of ODA-hectorite hybrid monolayer deposited onto ZnSe substrate at 15 mN/m along with the FTIR spectrum of hectorite and ODA in KBr pallet.

The ODA in KBr shows strong prominent bands at 2849 and 2919 cm$^{-1}$. These two bands are the diagnostic bands of ODA and identified as the stretching vibrations of $CH_2$ group of ODA.[27] These two bands are also present in the FTIR spectra of ODA-hectorite hybrid films. A slight shift of the band position is occurred due to the hybridization with the clay particles. The FTIR spectra of hectorite in KBr pallet shows prominent band at 1080 $cm^{-1}$, which is the out-of-plane Si–O stretching vibration of hectorite.[29] The in-plane Si–O stretching vibration (1010 cm$^{-1}$) is greatly reduced to a weak shoulder. The FTIR spectra of hybrid film on ZnSe substrate posses strong prominent bands at 1003 and 1078 cm$^{-1}$ due to the in-plane and out-of-plane Si-O stretching vibration respectively. Here a slight shift of the band position occurred due to the strong electrostatic interaction between the negatively charged clay and the positively charged ODA during hybrid monolayer formation. This confirms the presence of clay particles in the hybrid films.

### 3.3. *AFM investigation of hybrid monolayers*

The floating hybrid Langmuir monolayers were transferred onto smooth silicon substrates at surface pressures of 0 mN/m (before compression) and 15 mN/m$^{-1}$, and the morphology of the LB films were observed by AFM and shown in figure 4. The clay platelets are clearly observed in the images, which indicates the formation of the nanodimensional hybrid monolayer at the air-clay dispersion interface. Before the compression (0 mN m$^{-1}$, Figure 4a), there can be seen the isolated clay platelets and the empty space between them. But the AFM image of the hybrid LB films lifted at 15 mN/m (Figure 4b) shows dense and compact organization of clay platelets. Here the clay platelets are in contact with each other with very small vacant spaces in between them. The surface coverage is more than 80%. The elementary particles of hectorite are lath-like with a high aspect ratio. The film thickness is of the order of 2 nm. These thicknesses are the sum of the thickness of the elementary clay mineral platelet and the thickness of a monolayer of ODA molecules. Few overlapping aggregates of clay particles are also present in the films. These images confirmed that the hybridization of the ODA cations and the clay platelets (hectorite) were occurred before the compression and that the floating platelets were gathered as the film was compressed until they contacted each other at the onset of the surface pressure. By analyzing the AFM images (figure 4c) we tried to collect the information about the morphology of individual clay platelets. It is observed that the average height, length and width of individual clay platelets range between 1.5 to 2 nm, 500 to 1250 nm and 100 to 115 nm respectively.

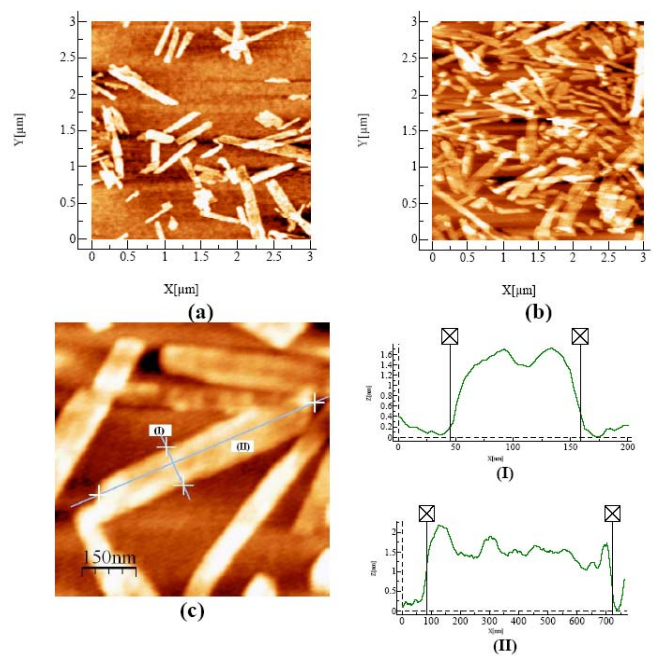

Fig. 4. AFM image of ODA-hectorite hybrid Langmuir monolayer deposited onto smooth silicon substrate deposited at (a) 0 mN/m ($3 \times 3 \mu m$), (b) 15 mN/m ($3 \times 3 \mu m$) and (c) 15 mN/m ($0.75 \times 0.75 \mu m$). Graph shows the line analysis for (I) width and (I) length with height profile.

## 4. Conclusions

In conclusion our results demonstrate the formation of hybrid monolayers of elementary clay mineral platelets hectorite and ODA cations by the LB technique. The monolayers consist of randomly oriented elementary clay mineral platelets of various sizes and shapes mixed with some aggregates. The elementary particles of hectorite are lath-like with a high aspect ratio. The monolayer thickness of hybrid film is 2 nm and average height, length and width of individual clay platelets ranges between 1.5 to 2 nm, 500 to 1250 nm and 100 to 115 nm respectively. The surface coverage was more than 80%. These thicknesses are the sum of the thickness of the elementary clay mineral platelet and the thickness of a monolayer of ODA molecules.

### Acknowledgement

The authors are grateful to CSIR, Govt. of India for providing financial assistance through CSIR project Ref. No. 03(1080)/06/EMR-II.

### References

1. G.H. Beall, in: L.L. Hench, S.W. Freeman (Eds.), *Advances in Nucleation and Crystallization in Glasses*, American Ceramics Society, 1971, p.5.
2. T.Hamasaki, K. Eguchi, Y. Koyanagi, A. Matsumoto, T. Utsunomiy, K.Kob, *J.Am.Chem. Soc*.**71** (1988) 1120.
3. T.J.Pinnavaia, J.-R.Butruille, in: G.Alberti, T.Bein (Eds.), *Comprehensive Supramolecular Chemistry*, vol.7, Elsevier, Oxford, 1996, Chapter 7.
4. T.J. Pinnavaia, G.W. Beall (Eds.), *Polymer–Clay Nanocomposites*, Wiley, New York, 2000.
5. H.Cr oss, S.Y ariv, *Organo-Clay Complexes and Interactions*, Marcel Dekker, 2001.
6. A.J.Bar d, T.Mallouk, in: R.W. Murray (Ed.), *Techniques of Chemistry Series*, vol.22, Wiley, New York, 1992, Chapter 6.
7. T.E. Mallouk, H.-N. Kim, P.J. Ollivier, S.W. Keller, in: G. Alberti, T.Bein (Eds.), *Comprehensive Supramolecular Chemistry*, vol.7, Elsevier, Oxford, 1996, Chapter 6.
8. R.A. Schoonheydt, in: G. Alberti, T.Bein (Eds.), *Comprehensive Supramolecular Chemistry*, vol.7, Elsevier, Oxford, 1996, Chapter 11.
9. M.Ogawa, K. Kuroda, *Chem.Rev* . **95** (1995) 399.
10. E.R. Kleinfeld, G.S. Ferguson, *Science* **265** (1994) 370.

11. I.Décány, T. Haraszti, *Colloids Surf.A* **123** (1997) 391.
12. N.A. Kotov, T.Haraszti, L.T uri, G. Zavala, R.E. Geer, I. Décány, J.H.Fendler, *J. Am.Chem. Soc.***119** (1997) 6821.
13. B. van Duffel, R.A. Schoonheydt, C.P.M. Grim, F.C. De Schryver, *Langmuir* **15** (1999) 7520.
14. B.van Duffel, T.V erbiest, S.V an Elshocht, A.Persoons, F.C. De Schryver, R.A. Schoonheydt, *Langmuir* **17** (2001) 1243.
15. Theng, B. K. G. *The Chemistry of Clay-Organic Reactions*; Adam Hilger: London, 1974.
16. Ogawa, M.; Kuroda, K. *Chem. Rev.* **95** (1995) 399.
17. Alberti, G.; Constantino, U. *In Comprehensive Supramolecular Chemistry*; Alberti, G., Bein, T., Eds.; Elsevier: Oxford, 1996; Chapter 1, Vol. 7
18. S. Jeong, W.H. Jang, J. Moon, *Thin Solid Films*, **466** (2004) 204-208.
19. J. Wang, H.S. Wang, L.S. Fu, F.Y. Liu, H.J. Zhang, *Mater. Sci. Eng. B*, **97** (2003) 83-86.
20. H.Y. Jing, X.L. Li, Y. Lu, Z.H. Mai, M. Li, *J. Phys. Chem. B*, **109** (2005) 2881-2884.
21. S.H. Gyepi-Garbrah, R. Silerova, *Phys. Chem. Chem. Phys.*, **4** (2002) 3436.
22. G.G. Roberts, *Langmuir–Blodgett Films*, Plenum Press, New York, 1990.
23. Laird, D. A. In CMS Workshop Lectures; Mermut, A. R., Ed.; The Clay Minerals Society, Aurora, CO, 1994; Vol. 6, p 79.
24. Kawamata, J.; Ogata, Y.; Taniguchi, M.; Yamagishi, A.; Inoue, K. *Mol. Cryst. Liq. Cryst.* **343** (2000) 53.
25. M.A. Kalinina, V.V. Arslanov, S.I. Zheludeva, E.Y. Tereschenko, *Thin Solid Films*, **472** (2005) 232.
26. T. Tsukatani, H. Fujihara, *Langmuir*, **21** (2005) 12093.
27. Robin H. A. Ras, Cliff T. Johnston, Robert A. Schoonheydt *Chem. Commun.*, (2005) 4095
28. Gaines Jr. G. L. *Insoluble Monolayers at Liquid-Gas Interface*, John Wiley & Sons, New York, 1966.
29. J. Madejová, J. Bujdák, M. Janek, P. Komadel *Spectrochim. Acta A* **54** (1998) 1397.